\documentclass[aps,prb,signlecolumn,preprint,superscriptaddress,10]{revtex4-1}
\usepackage{hyperref	}
\usepackage{graphicx}
\usepackage{bm}
\usepackage{subfigure}

\usepackage{textcomp} 
\usepackage{amssymb}
\usepackage{amsmath}
\usepackage{mathptmx}
\usepackage{xspace}
\usepackage{mathrsfs}
\usepackage[usenames,dvipsnames]{color}

\newcommand{\HH}{\ensuremath{\mathscr{H}}}
\newcommand{\smmu}{\ensuremath{\mu_{\mathrm{s}}}\xspace}
\newcommand{\smJij}{\ensuremath{J_{ij}}\xspace}

\newcommand{\smKu}{\ensuremath{k_{\mathrm{u}}}\xspace}

\newcommand{\sms}{\ensuremath{\mathbf{S}}\xspace}

\newcommand{\vampire}{\textsc{vampire}\xspace}

\newcommand{\kB}{\ensuremath{k_{\mathrm{B}}}\xspace}
\newcommand{\muB}{\ensuremath{\mu_{\mathrm{B}}}\xspace}
\newcommand{\Mags}{\ensuremath{M_{\mathrm{s}}}\xspace}

\newcommand{\Keff}{\ensuremath{K_{\mathrm{eff}}}\xspace}
\newcommand{\As}{\ensuremath{A_{\mathrm{s}}}\xspace}

\newcommand{\Ang}{\textup{\AA}\xspace} 

\begin{document}
\title{Thermally nucleated magnetic reversal in CoFeB/MgO nanodots}
\author{Andrea Meo}
\email{am1808@york.ac.uk}
\affiliation{Department of Physics, University of York, Heslington, York YO10 5DD United Kingdom.}
\author{Phanwadee Chureemart}
\affiliation{Computational and experimental magnetism group, Department of Physics, Mahasarakham University, Mahasarakham, Thailand}
\author{Shuxia Wang}
\affiliation{Samsung Electronics, Semiconductor R\&D Center (Grandis), San Jose, CA 95134, USA}
\author{Roman Chepulskyy}
\affiliation{Samsung Electronics, Semiconductor R\&D Center (Grandis), San Jose, CA 95134, USA}
\author{Dmytro Apalkov}
\affiliation{Samsung Electronics, Semiconductor R\&D Center (Grandis), San Jose, CA 95134, USA}
\author{Roy W. Chantrell}
\affiliation{Department of Physics, University of York, Heslington, York YO10 5DD United Kingdom.}
\author{Richard F. L. Evans}
\affiliation{Department of Physics, University of York, Heslington, York YO10 5DD United Kingdom.}

\begin{abstract}
Power consumption is the main limitation in the development of new high performance random access memory for portable electronic devices. Magnetic RAM (MRAM) with CoFeB/MgO based magnetic tunnel junctions (MTJs) is a promising candidate for reducing the power consumption given its non-volatile nature while achieveing high performance. The dynamic properties and switching mechanisms of MTJs are critical to understanding device operation and to enable scaling of devices below 30 nm in diameter.
Here we show that the magnetic reversal mechanism is incoherent and that the switching is thermally nucleated  at device operating temperatures. Moreover, we find an intrinsic thermal switching field distribution arising on the sub-nanosecond timescale even in the absence of size and anisotropy distributions or material defects. 
These features represent the characteristic signature of the dynamic properties in MTJs and give an intrinsic limit to reversal reliability in small magnetic nanodevices. 
\end{abstract}

\maketitle
Recent advances in low power computing technology have enabled the development of high performance portable computing devices such as smart phones and tablet computers. A limiting factor today for mobile and high performance systems is the power consumed by the main system memory which is based on volatile Dynamic Random Access Memory (DRAM). The volatility arises due to electron leakage, requiring frequent refreshing of the stored data resulting in the memory consuming between 30\% and 50\% of the total system power\cite{VetterCSE2015}. Magnetic RAM (MRAM) is a non-volatile solid state memory technology~\cite{MRAM_1200123} based on a magnetic tunnel junction (MTJ) where the data are stored as a magnetic state rather than electrical charge~\cite{Gajek2012,samsung-review,7428818}. The non-volatile nature of the data removes the need for refreshing the data leading to a large reduction in power consumption as well as higher performance. 

CoFeB/MgO/CoFeB MTJs have attracted particular interest due to their high thermal stability, low damping and high tunnel magneto resistance (TMR). 
Thermal stability is determined by the magnetic anisotropy of the device, which in CoFeB MTJs arises due to hybridisation of the atomic orbitals of the magnetic layer and the MgO interface~\cite{samsung-review,:/content/aip/journal/apl/105/22/10.1063/1.4903296,Peng2015}. In CoFeB/MgO the anisotropy is sufficient to provide thermal stability and to support an out-of-plane magnetization. High TMR is achieved because MgO acts as a good spin filtering barrier and the good crystallisation of both CoFeB and MgO preserves the spin polarisation of the electrons crossing the MTJ~\cite{TMR_CoFeMgO,:/content/aip/journal/jap/101/1/10.1063/1.2407270,4160113,samsung-review,CoFeB-MgO_TMR}. Damping is low due to the weak spin-orbit coupling characterising CoFe-alloys and the good crystalline quality of the film, which is required to reduce the critical current for spin transfer torque (STT) switching~\cite{samsung-review}.

Despite the promising intrinsic properties of CoFeB/MgO, patterned nanoscale devices introduce many complexities including finite size and surface effects, strong magnetostatic interactions and complex magnetization dynamics. Previous experimental~\cite{:/content/aip/journal/apl/107/15/10.1063/1.4933256,Sato2014,6189858} and micromagnetic studies~\cite{Jang2015,Sampaio2016,Song2015} 
 have concluded that the reversal mechanism is likely to be incoherent due to the large lateral size of the devices. However, the nature of the reversal mechanism and in particular the effects of the localised anisotropy induced at the CoFeB/MgO interface and of the temperature are currently unknown. In addition the role of magnetostatic coupling between the free and reference layers of an MTJ device is an open question due to the strength of the interactions caused by their proximity. 
A conventional micromagnetic model approaches the limit of validity at sizes relevant for technological applications of MTJs. The discretisation of the system into micromagnetic cells and the fact that the minimum cell size is around $1\,\mathrm{nm}^3$ precludes the possibility of taking into account the atomic variation of properties which occurs in these systems whose thickness is of the order of few nanometers, e.g. the fact that the anisotropy is localised at the atomic interface between CoFeB and MgO. In addition, finite temperature effects are poorly described because atomic spin fluctuations are neglected and finite size effects that play an important role in determining the thermal stability of the system for in-plane dimensions below 50~nm, cannot be properly captured. The presence of interfaces causes the reduction of surface coordination and hence loss of exchange bonds at the surface, which leads to lower exchange coupling than in a bulk system. The micromagnetic approach tends to underestimate this effect and often only the dynamics of the free layer is considered.
Determining the reversal mechanism is critical in evaluating the thermal stability and the switching time in spin transfer torque MRAM devices. We aim to investigate the dynamics at the atomistic level not constrained by the limits of micromagnetic models. Using an atomistic spin model, we simulate the dynamic properties of CoFeB/MgO nanodots and MTJs. The results of the simulations demonstrate two main features: 1) the magnetization reversal is incoherent for in-plane dimensions larger than 30~nm; and 2) the switching of the magnetization is thermally driven at temperatures at which devices operate. A domain wall is nucleated at the edge of the system which then propagates through the disk, leading to coercive fields lower than for the case of coherent reversal. The fact that the switching is thermally driven poses an intrinsic limitation to the deterministic reversal process and reduces the thermal stability for small devices.  
\section*{Results}
\subsection*{Atomistic modelling of field induced magnetization reversal}
\begin{figure}[!tb]
    \centering
		\includegraphics[width=0.5\columnwidth]{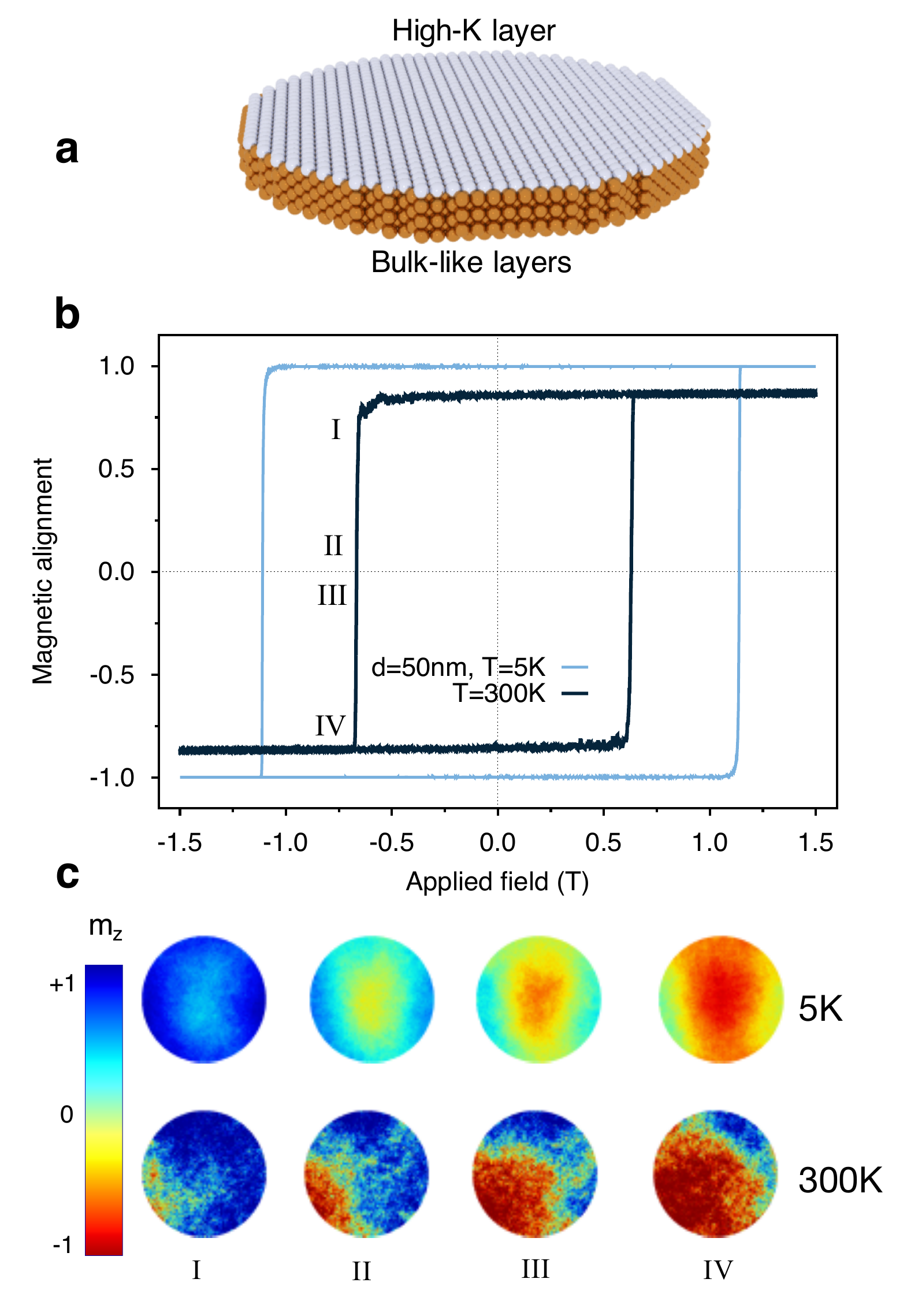}
			\caption{(a) Schematic of the simulated system with light spheres representing the high anisotropy layer, and dark spheres representing the bulk-like CoFeB layer. (b) Typical simulated easy-axis hysteresis loop for 1~nm thick, 50 nm diameter nanodot at temperatures of 5 K and 300K. The data show a large reduction in the coercivity for elevated temperatures due to increased thermal fluctuations, indicating a change in  the magnetic reversal mechanism. (c) Snapshots of magnetization reversal at 5 and 300~K for disk of diameter 50~nm and thickness 1~nm. I and IV refer to the top and bottom shoulder of M/\Mags vs  H curve, respectively. II and III are configurations just before and after the switching, respectively. The color scheme represents the magnetization along the easy axis direction (z).}
		\label{fig:dots}
\end{figure}

Using an atomistic spin model based on Heisenberg Hamiltonian as implemented in the \vampire software package~\cite{vampire,vampire-rev} we simulate the dynamic properties of CoFeB/MgO nanodots, a schematic of which is presented in Fig.~\ref{fig:dots}(a). Thermal effects  are included in the model via a Gaussian white noise term whose amplitude is temperature dependent and the magnetostatic contribution a macrocell approach. Details of the model are given in the methods section. We focus on the hysteretic properties of CoFeB/MgO nanodots, in particular investigating the temperature, size and thickness dependence of the reversal mechanism and dynamic coercivity. After investigating the field-driven dynamics we conclude with simulations of spin torque switching.
\subsection*{Thermal effects}
We first consider the effects of temperature on the typical hysteresis properties of a nanodot with a diameter of 50 nm, as shown in Fig.~\ref{fig:dots}(b). The first observation is that increased temperatures lead to a large reduction in the coercivity from 1.1~T at 5~K to 0.6~T at 300~K. The temperature variation of intrinsic properties such as the saturation magnetization and magnetic anisotropy arises naturally from the atomistic simulations, using Monte Carlo methods as outlined in the methods section. This leads to an expected 20\% reduction in $H_K$ between zero and 300K
but here we a observe a 45\% reduction in the coercivity. This is partially due to the thermally activated transitions over the energy barrier, but also may reflect  a change in the magnetic reversal mechanism due to the stronger thermal fluctuations. To investigate the reversal mechanisms we have generated snapshots of the atomic spin configuration during hysteresis for different temperatures, as shown in Fig.~\ref{fig:dots}(c). At a temperature of 5~K the reversal is semi-coherent and nucleated at the centre of the nanodot due to the larger magnetostatic field. At 300 K the reversal is initiated by the nucleation of a small reversed domain at the edge of the nanodot caused by thermally driven spin fluctuations at the edge. At the edge a loss of exchange bonds leads to larger edge spin fluctuations compared with the spins in the middle of the dot. These larger spin fluctuations provide a natural nucleation region at the edges of the nanodot and therefore allow a different reversal mechanism compared to the centre nucleated reversal at low temperatures.
Interestingly the small size of the system means that the thermal fluctuations are more important than the variation in the magnetostatic field across the dot diameter, highlighting the importance of including thermal fluctuations and surface effects in the model compared with non-stochastic continuum micromagnetic simulations.

We note that the thermally nucleated switching we describe here is different from the Sharrock approach~\cite{sharrock} which considers a fixed (coherent) reversal mechanism but with a time dependence of the magnetization due thermally  induced transitions over the energy barrier. In the case of CoFeB/MgO dots the thermal fluctuations lead to a large reduction in the coercivity due to the ability to access a different thermally driven reversal mode. Of course, slower hysteresis loops will likely lead to a further reduction in the coercivity in a similar manner to that of Sharrock due to the increased number of nucleation attempts, but such simulations are currently beyond the timescales accessible with atomistic models. 

Another interesting feature of the hysteresis loop at 300~K in Fig.~\ref{fig:dots}(b) is a slight asymmetry in the coercivity of the ascending and descending branches of the loop. This is due to the thermally nucleated nature of the reversal, leading to an uncertainty in the exact coercivity due to the randomness of the nucleation attempts. There is therefore an intrinsic thermal switching field distribution which is independent of defects and variations in the intrinsic properties, but arises solely due to random thermal fluctuations. For larger systems and long timescales the thermal switching field distribution is not apparent, but for nanoscale MTJs switching in the nanosecond time domain it is a real and important effect and represents the thermodynamic limit of the switching field distribution which cannot be overcome. 

\subsection*{Size effects}
\begin{figure}[!tb]
	\centering	\includegraphics[width=0.5\columnwidth]{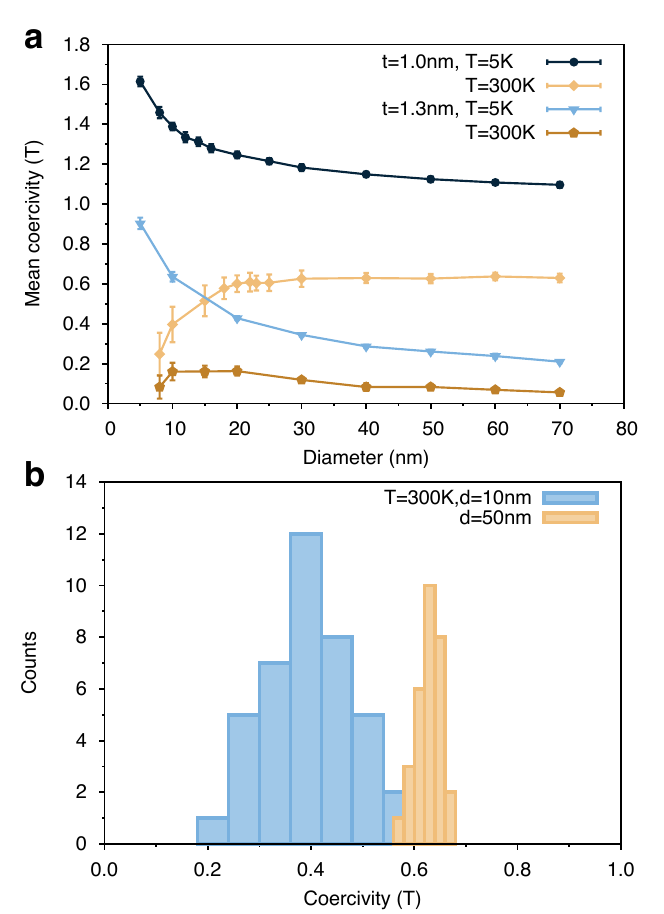}	
	\caption{(a) Mean coercivity as function of disk diameter for thicknesses of 1.0 and 1.3~nm at 5 and 300~K. Error bars show the standard deviation of the statistical distribution. (b) Calculated switching field distributions at 300~K for 10 and 50~nm nanodots.}
	\label{fig:SFD}
\end{figure}

To investigate the effects of nanodot size and temperature on the coercivity and thermal switching field distribution we have performed a systematic study of the hysteretic properties for 1~nm and 1.3~nm thick nanodots, shown in Fig.~\ref{fig:SFD}(a). The size dependence of the coercivity is obtained by averaging over a minimum of 30 independent loops for each size, temperature and thickness.
The mean coercivity shows a complex temperature and size dependence  which is due to different reversal mechanisms and finite effects. Considering first the 1~nm thick nanodots, the coercivity reaches an asymptotic limit for nanodot diameters $> 20$~nm indicative of a nucleation reversal mode at  300~K with a slower approach at 5~K. However the snapshots of the atomic spin configurations support the earlier conclusion of different reversal modes at low and room temperature respectively. At 5~K the nucleation is driven by the variation of the magnetostatic field across the nanodot, which increases with increasing nanodot diameter leading to a slow convergence to a constant nucleation field only seen for larger nanodot diameters (around 100~nm, Supplementary Fig.3, Supplementary Note 2). Conversely at 300~K the thermal nucleation volume is much smaller and independent of the dot size, and so the coercivity reaches an asymptotic limit at around 20~nm diameter. For dots smaller than 20~nm diameter the temperature has a dramatic effect on the coercivity, showing a large increase at 5~K and large decrease at 300~K respectively. We note that the increase of coercivity with decreasing diameter is indicative that the system has not reached the critical diameter for superparamagnetic behavior.

 For low temperatures the increase in the coercivity with decreasing diameter indicates a transition to coherent reversal (see Supplementary Fig.2, Supplementary Note 1), where the magnetostatic field no longer dominates the reversal process and the nanodot size approaches the single domain limit $\delta_w = \pi \sqrt{\As/\Keff} \sim 10$~nm. At room temperature the reduction in the coercivity is due to superparamagnetic fluctuations of the magnetization which due to the small volume lead to switching at fields lower than the intrinsic coercivity.
 We note here that unlike the work of Brown~\cite{Brown1963} there is no peak in the coercivity due to the approximately two-dimensional nature of the nanodots and large anisotropy, hence the direct transition between superparamagnetic and nucleated reversal behaviour as a function of the nanodot size. For dots of diameter smaller than 10~nm the system enters in a single domain limit (in agreement with the estimation of the single domain size $\delta_w$) and at room temperature the system becomes unstable due to a transition towards SPM regime. 
 The 1.3~nm thick nanodots show a similar qualitative behavior as the 1~nm thick nanodots as a function of the nanodot size, though with a significantly reduced coercivity. The large reduction in the coercivity arises mainly from the reduced anisotropy because of its proportionality with 1 /thickness energy. In addition, a change in the magnetostatic energy caused by the increased thickness contributes to the coercivity decrease.
The combination of these effects reduces the stability of the perpendicular orientation of the magnetization and therefore increases the stability of nucleated domains under an applied field. 

The statistical distribution of the  coercivity for different nanodot sizes and temperatures is also strongly size dependent. The extracted switching field distributions (SFD) at room temperature for diameters of 10 and 50~nm and thickness 1~nm are presented in Fig.~\ref{fig:SFD}(b). The distributions show a range of switching fields which is much larger for the smaller nanodot size. In the case of our simulations, each nanodot of a given size is identical in terms of the number of atoms and magnetic parameters, but with different pseudorandom number sequence representing the random nature of the thermal noise in the sLLG equation. Therefore the origin of this distribution is purely the random thermal fluctuations during the reversal process, and hence the distribution is the \emph{thermal} switching field distribution (TSFD)~\cite{Breth2012}. At the switching field the time scale of the reversal is determined by these random thermal fluctuations, leading to a natural TSFD for a switching process on the timescale of a few nanoseconds. The TSFD is an intrinsic property of small magnetic elements and cannot be overcome due to its intrinsic thermodynamic origin. We note that the TSFD is also thickness dependent, being narrower for thicker films due to the reduced thermal fluctuations associated with the larger magnetization volume. Importantly the TSFD intrinsically limits the ability to reliably reverse a nanodot at a given field and timescale, leading to a natural distribution of switching probability for a finite time and strength field pulse \cite{PhysRevB.94.014404}. 

\subsection*{MTJ switching}
\begin{figure}[!tb]
	\centering
	\includegraphics[width=0.5\columnwidth]{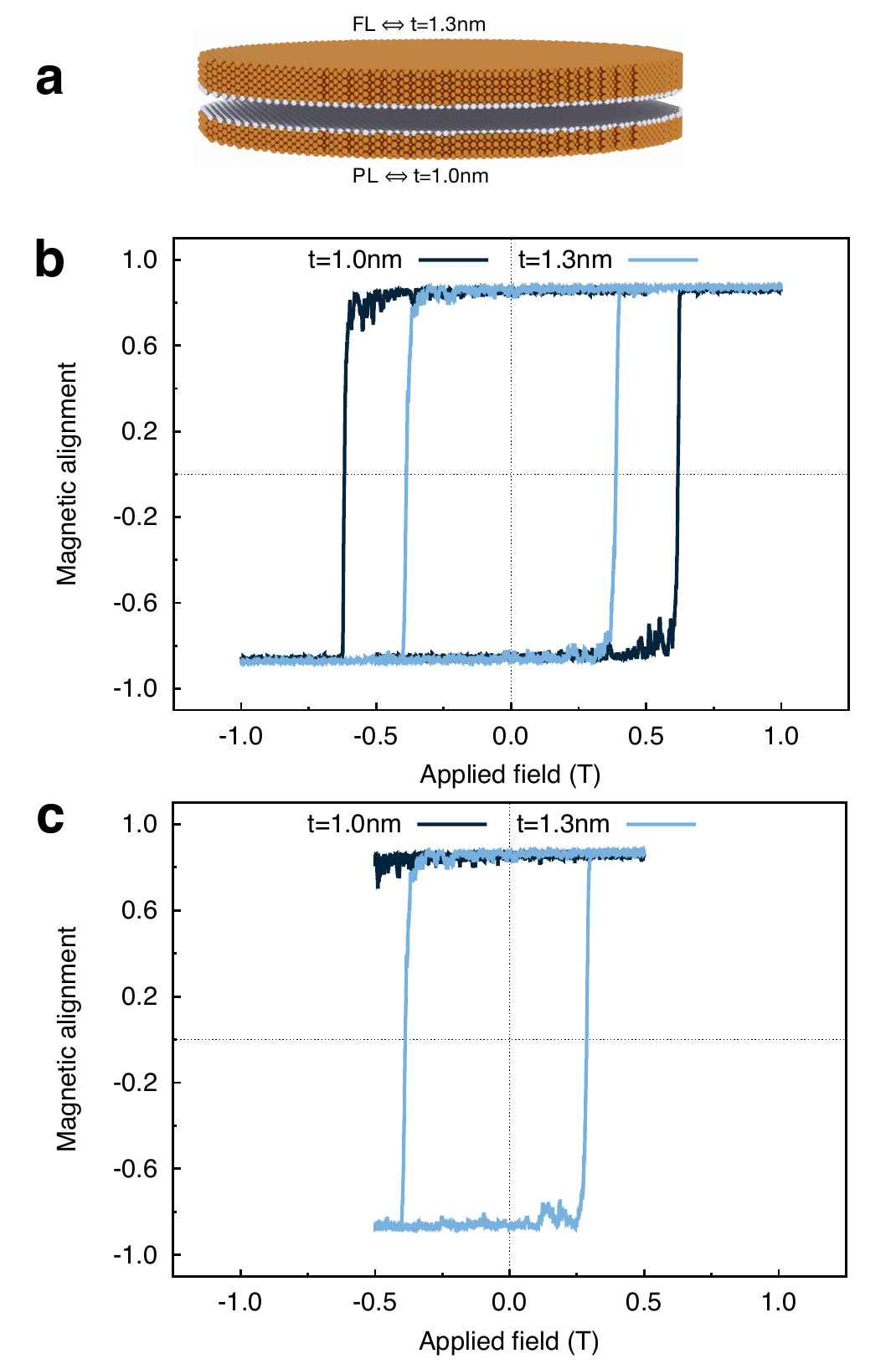}
	\caption{(a) Schematic of the simulated MTJ structure. Major (b) and minor (c) hysteresis loops for an MTJ of diameter 30~nm at 300~K. The major loops show a large enhancement of both layer coercivities due to the coupling to the stray field. The minor loop exhibits a shift of the hysteresis loop due to the asymmetric effect of the pinned layer stray field for descending and ascending branches.}
	\label{fig:MTJ_minorloop}
\end{figure}

\begin{figure}[!tb]
	\centering
	\includegraphics[width=0.5\columnwidth]{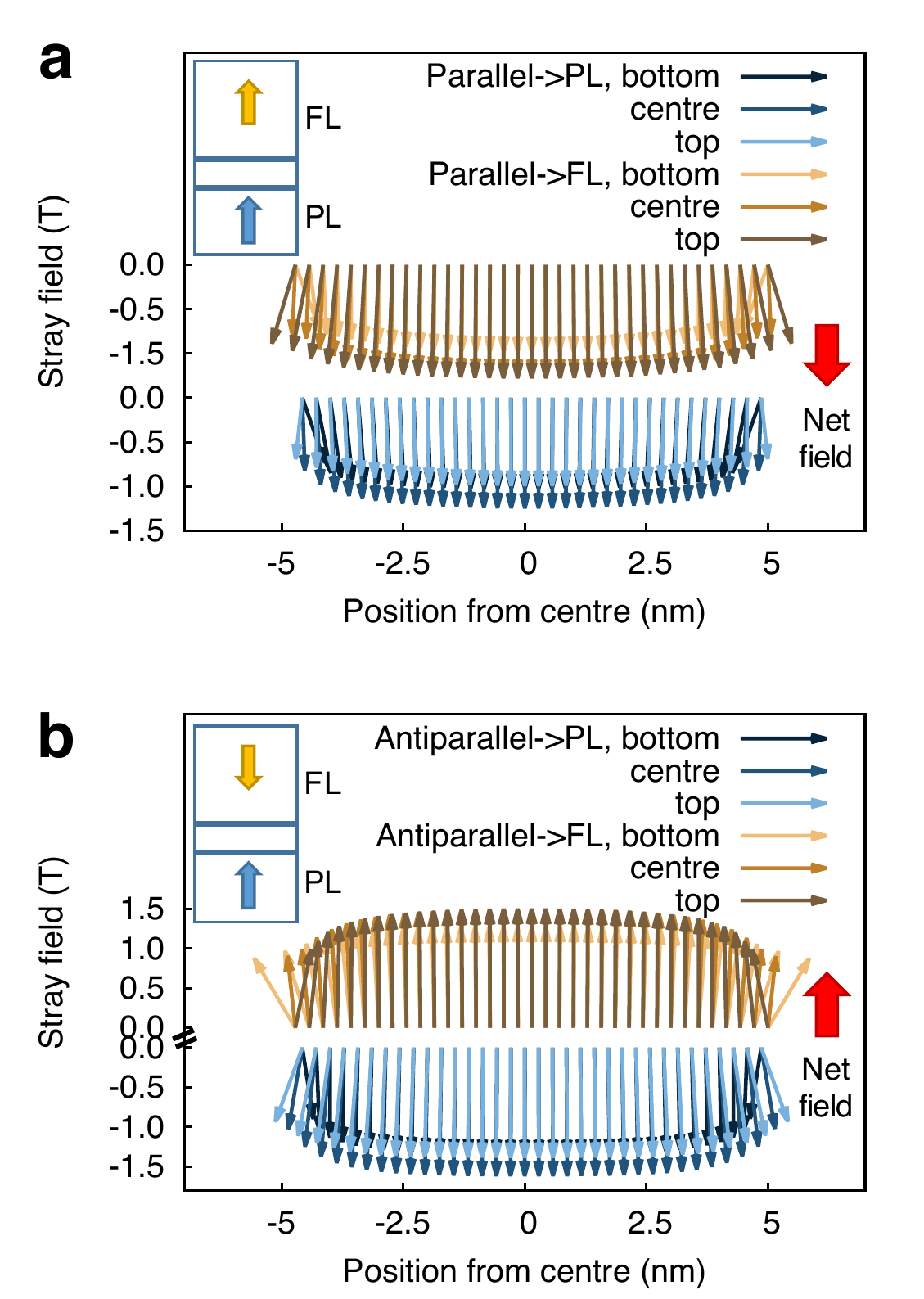}
	\caption{Stray field generated by three atomic layers in PL and FL in parallel (a) and anti-parallel (b) configuration as function of position from the center of the disk. The insets show a schematic of the layer magnetization and the net average stray field.}
	\label{fig:MTJ_strayfield}
\end{figure}

So far we have considered the properties of isolated CoFeB/MgO nanodots, however the close proximity of the layers in an MTJ device leads to a significant magnetostatic interaction between the layers. We have investigated the dynamics and magnetization reversal, including the effects of magnetostatic interactions, in an MTJ structure with dimensions CoFeB(1.0nm)[PL]/ MgO(0.85nm)/ CoFeB(1.3nm)[FL] and 30~nm diameter, shown schematically in Fig.~\ref{fig:MTJ_minorloop}(a). 
Due to the strong coupling in MTJ, we have modified the usual macrocell approach for the calculation of the magnetostatic field following the approach proposed by Bowden in~\cite{bowden} to obtain exact agreement with the atomic scale dipole-dipole interaction assuming a uniform magnetization in each cell, a good approximation for our cell size of 1~nm. We have calculated major and minor hysteresis loops for the MTJ structure at room temperature as shown in Fig.~\ref{fig:MTJ_minorloop}(b) and (c) respectively. We find that the free and reference layers switch independently and that the reversal mechanism exhibits the same features observed for the individual layers, that of thermally nucleated switching (see Supplementary Fig.4-5, Supplementary Note 3). In major loops, compared to the single layer coercivities the magnetostatic coupling in the MTJ tends to stabilize the magnetic structure and enhances the coercivity of both layers compared to the free nanodots. In the minor loop, shown in Fig.~\ref{fig:MTJ_minorloop}(c), the free layer exhibits a bias due to the stabilizing (destabilizing) effect of the magnetostatic field from the pinned layer for the descending (ascending) branches. 
To quantify the  the magnetostatic field from the pinned layer acting on the free layer we have calculated the stray field with atomic resolution as function of position and the  net average stray field 
in Fig.~\ref{fig:MTJ_strayfield}, showing the existence of a stabilizing (destabilizing) field depending on magnetic configuration.
In the case of MTJ devices the strong coupling of the magnetic layers leads to a a complex change in the magnetic properties such as coercivity.
We have also investigated the effect of thermal fluctuations on spin transfer torque switching mechanism following Slonczewski's approach~\cite{Slonczewski1996}. 
\begin{figure}[!tb]
	\centering
	\includegraphics[width=0.5\columnwidth]{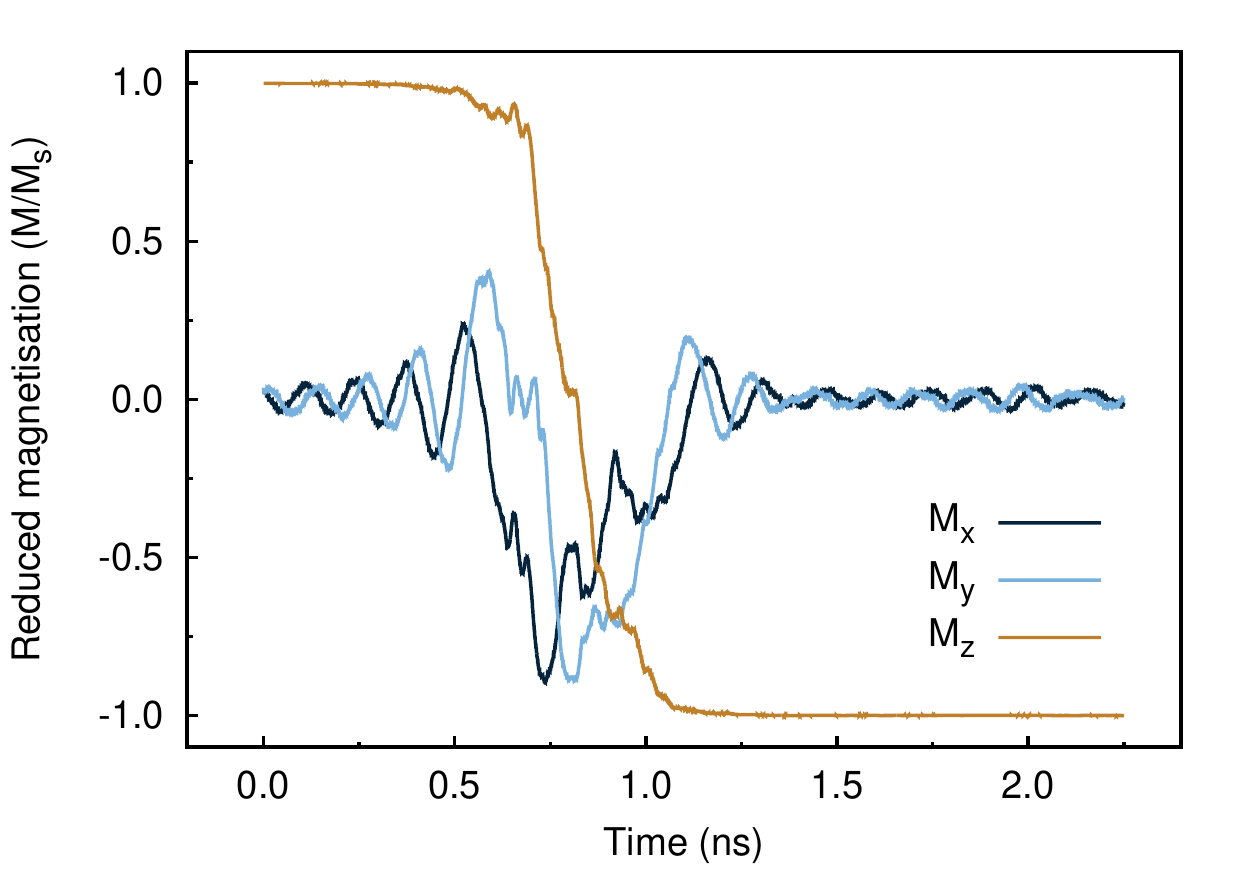}
	\caption{Time evolution of magnetization at room temperature of a 40~nm diameter MTJ. The colours refer to the different component of the magnetization.}
	\label{fig:STT-time-evol}
\end{figure}
Fig.~\ref{fig:STT-time-evol} show the time evolution of magnetization for a MTJ of diameter 40~nm at room temperature. We observe that the magnetization is reversed in the order of a nanosecond, in agreement with switching times measured experimentally by Devolder \textit{et al} in~\cite{Devolderjan2016}. From the analysis of the spin configurations during the spin transfer torque switching (see Supplementary Note 4), a thermally activated incoherent reversal occurring via domain nucleation at the edge of the dot emerges. This result is in agreement with the reversal mechanism induced by an applied field and agrees with Devolder \textit{et al}~\cite{DevolderJune2016} who find that the spin transfer torque switching is thermally activated and characterized by domain wall propagation for comparable in-plane MTJ dimensions, therefore confirming the thermal nature of the switching and the resulting non-collinear character in such systems.

\section*{Discussion}
We have investigated the magnetization reversal mechanism in CoFeB/MgO nanodots and MTJs using an atomistic spin model with the inclusion of thermal and magnetostatic effects.
The magnetisation reversal in CoFeB/MgO nanodots and MTJs can be described as thermally nucleated and incoherent at temperatures relevant to device operation, leading to a large reduction in the coercivity compared to a coherent reversal mechanism. The thermal nature of the reversal mechanism is also reflected in  the spin transfer torque switching mechanism of MTJ devices, hence affecting the reversal speed. In an MTJ geometry we find that the magnetostatic interaction between the layers leads to a stabilizing effect on both the pinned and free layers and causes a shift of the minor hysteresis loop.  
Our results highlight the importance of considering at an atomistic level finite size and thermal fluctuations when modelling such small scale magnetic devices which can have a dominant effect on their reversal mechanisms and physical properties. It is important to note the large difference between the coercivities in our simulation of a perfect nanodot and those measured experimentally, where coercivities are typically $\sim 0.1$~T. In our model we have used material parameters derived from experimental measurements of continuous thin films, and as such our simulations represent the best case situation concerning large coercivity. We expect that realistic devices are affected by edge damage and defects which lead to a further reduction of the coercivity. Our results also raise further questions on the role of thermal fluctuations on spin transfer torque switching and the energy barrier in zero field responsible for the thermal stability of MTJs. We expect that the timescale of the spin transfer torque switching is strongly dependent on the lateral size of the MTJ due to different magnitudes of the thermal fluctuations breaking the magnetic symmetry required for switching and will be the subject of future work.

\section*{Methods}
\subsection*{Atomistic spin model}
The simulations were performed using an atomistic spin model where the energy of the system is described by a classical Heisenberg Hamiltonian  (\HH)
\begin{equation}
\HH = -\sum_{i\ne j} \smJij \sms_i \cdot \sms_j -\sum_{i} \smKu S_{i,z}^2 - \sum_i \smmu \sms_i \cdot \mathrm{\mathbf{H}_{app}} + \HH_{\mathrm{demag}}.
\label{eq:Heisenberg-Hamiltonian}
\end{equation}
where $\sms_{i,j}$ are normalized spin vectors on site i,j respectively, \smJij is the exchange  coupling between spin i and j, \smKu is the single-ion uniaxial magnetocrystalline anisotropy energy (MAE) constant per site, \smmu the atomic spin moment,  $\mathrm{\mathbf{H}_{app}}$ the applied external field and $\HH_{demag}$ the magnetostatic contribution. First and second term on RHS  of Eq.~\ref{eq:Heisenberg-Hamiltonian} describe a system with nearest neighbours isotropic exchange interactions and uniaxial MAE respectively, while the third term represents Zeeman's interaction with an external field~\cite{vampire-rev}.  Given the high computational cost required to calculate calculate the magnetostatic energy due to the long range nature of this interaction, the demagnetization field is computed applying a micromagnetic discretization of the system into macrocells that are considered as dipoles. Each macrocell \textit{i} has a magnetic moment $\mathrm{\mathbf{m}}_i^{\mathrm{mc}}$ determined by the vector sum of the atomic spin moments inside the cell and position calculated from the magnetic centre of mass of the cell and volume $V_p^{\mathrm{mc}}$~\cite{vampire-rev}. The magnetostatic energy 
$\HH_{demag}$ takes the form $-\frac{1}{2}\sum_p \mathrm{\mathbf{m}}_p^{\mathrm{mc}} \cdot \mathrm{\mathbf{H}}^p_{\mathrm{demag}}$ with $\mathrm{\mathbf{H}}^p_{\mathrm{demag}}$ the magnetostatic field within the macrocell given by:
\begin{equation}
\mathrm{\mathbf{H}}^p_{\mathrm{demag}} = \frac{\mu_0}{4 \pi} 
\left(
\sum_{p\neq q}\frac{3 \left( \mathbf{m}_q^{\mathrm{mc}} \cdot \mathbf{\hat{r}}\right) \mathbf{\hat{r}} - \mathbf{m}_q^{\mathrm{mc}} }{r^3} 
\right) - \frac{\mu_0}{3}\frac{ \mathbf{m}_p^{\mathrm{mc}} }{ V_p^{\mathrm{mc}} }
\label{eq:demag_field}
\end{equation}
where \textit{r} is the distance between macrocells \textit{p}, \textit{q} and $\mathbf{\hat{r}}$ is a unit vector pointing along the direction $\overrightarrow{pq}$.  In equation~\ref{eq:demag_field} the first term represent dipolar field acting on macrocell \textit{p} due to all the other macrocells, the second accounts for the self-demagnetization field experienced by the moment of the macrocell $\mathrm{\mathbf{m}}_p^{\mathrm{mc}}$ itself. It is important to note that this approach requires the size of the macrocell used to discretize the system to be much smaller than the system size.

The dynamics of magnetization of CoFeB/MgO nandots and MTJs is determined solving the Landau-Lifshitz-Gilbert equation of motion~\cite{vampire-rev}, given by:
\begin{equation}
\frac {d \sms_i}{dt} =  - \frac{\gamma}{\left( 1+\alpha^2 \right)}\left[ \sms_i \times \mathrm{\mathbf{H}^i_{eff}} + \alpha \sms_i \times \left( \sms_i \times \mathrm{\mathbf{H}^i_{eff}} \right)\right].
\end{equation}
$\gamma$ is the gyromagnetic ratio of the electron, $\alpha$ the Gilbert damping which describes the relaxation of the atomic spins caused by electron-electron and electron-lattice interactions, $\sms_i$ is the unitary spin vector on site $i$ and $\mathrm{\mathbf{H}^i_{eff}}$ is the effective field acting on the spin $i$. The simulations are performed in a critical damping regime where $\alpha=1$ in order to allow a faster relaxation of the magnetization along the direction of $\mathrm{\mathbf{H}^i_{eff}}$, while the mechanism is not affected. $\mathrm{\mathbf{H}^i_{eff}}$ is obtained differentiating the Hamiltonian~\ref{eq:Heisenberg-Hamiltonian} with respect to $\sms_i$. The effect of temperature is introduced by adding a white noise term to $\mathrm{\mathbf{H}_{eff}}$ given the uncorrelated nature of thermal fluctuations on the considered time-scale ($\geq$ ns) following the approach proposed by Brown~\cite{Brown1979}. The thermal field $\mathrm{\mathbf{H}^i_{th}}$ is expressed as:
\begin{equation}
\mathrm{\mathbf{H}^i_{th}} =  \mathbf{G}\left(t \right)\sqrt{\frac{2\alpha\kB T}{\gamma \smmu \Delta t}}
\end{equation}
where $\mathbf{G}\left(t \right)$ is a Gaussian distribution in three dimensions, \kB is the Boltzmann constant, $\alpha$ the Gilbert damping, $\gamma$ the gyromagnetic ratio, $T$ the temperature, $\Delta t$ the time step used to integrate the equation of motion and \smmu the atomic is spin moment. The stochastic LLG equation of motion is solved by means of a Heun predictor-corrector algorithm, particularly suitable to deal with stochastic phenomena~\cite{vampire-rev}. 
The spin transfer torque contribution to the field is included in the LLG dynamics based on the work of Slonczewski~\cite{Slonczewski1996} and Fert \textit{et al}~\cite{Zhang2002} by adding to the effective field the term:
\begin{equation}
\mathrm{STT} = a \left(\sms_i \times \mathrm{\mathbf{M}_{p}}\right) + b \mathrm{\mathbf{M}_{p}}
\label{eq:STT_slonc}
\end{equation}
where $\sms_i$ is the unitarian spin vector on site \textit{i}, $\mathrm{\mathbf{M}_{p}}$ is the unit vector describing the direction of the injected current and \textit{a,b} are the adiabatic and non-adiabatic spin torque parameters which depend on the applied current density and material properties. As \textit{a} and \textit{b} we extract the values from ~\cite{Zhang2002} that  correspond to a similar spin-valve structure and current density of $1\times 10^{11} \mathrm{A}\,\mathrm{m}^{-2}$.
The temperature dependence of static magnetic properties $M(T)$ and $K(T)$ were calculated using respectively conventional Monte Carlo methods and the Constrained Monte Carlo approach~\cite{CMC_vampire}
\subsection*{Investigated system}
We consider an idealized model where all of the magnetic anisotropy is provided by a single monolayer of CoFeB in contact with the non-magnetic MgO and the other layers contribute no anisotropy. The elemental properties of Fe, Co and B are not considered, but treated as an average magnetic material with zero anisotropy. The atomic structure of CoFeB is modelled as a bcc lattice with lattice constant 2.86~\Ang and the bulk bcc crystal is cut into the shape of a cylinder of thickness 1.0 and 1.3~nm, representing the reference layer (RL) and the free layer (FL) for the MTJ respectively as shown in Fig.~\ref{fig:dots}(a).
The non-magnetic MgO oxide layer is not included in the simulations explicitly. \textit{Ab-initio} studies~\cite{Yang2011,Turek2003} suggest that MgO induces a strong interfacial perpendicular anisotropy at the interface CoFeB/MgO and enhances the exchange coupling of Fe and Co sites at the same interface, therefore we model these properties using effective anisotropy and exchange parameters obtained from direct comparison with experiments~\cite{sato2016}.  
The atomic spin moment used for our simulations is $\smmu = 1.60~\muB$ corresponding to $\Mags \sim 1.3\, \mathrm{MA}\,\mathrm{m}^{-1}$, close to the experimental value~\cite{ikeda}. The value of the atomic spin moment in our simulations is significantly lower than expected experimentally for bulk CoFe or from \textit{ab-initio} calculations of CoFe/MgO, where values close to $2.5~\muB$ are found~\cite{:/content/aip/journal/apl/90/8/10.1063/1.2710181,Bose2016}. Experimentally the CoFeB/MgO system is known to have a perpendicular orientation for effective thicknesses less than 1.2 - 1.3~nm~\cite{ikeda,:/content/aip/journal/apl/107/15/10.1063/1.4933256,C5NR01140J,:/content/aip/journal/adva/2/4/10.1063/1.4771996,PhysRevB.90.184409,:/content/aip/journal/apl/102/24/10.1063/1.4811269,:/content/aip/journal/apl/105/22/10.1063/1.4903296,:/content/aip/journal/apl/106/26/10.1063/1.4923272} and hysteresis simulations for different atomic moments (Supplementary Fig.1, Supplementary Methods) confirm that an effective atomic moment less than $2~\muB$ is required to have perpendicular orientation of the magnetization and square loops. The physical origin of the reduced saturation magnetization is likely due to a combination of the presence of non-magnetic Boron and the possibility of structural defects in the material. 
The effect of the demagnetizing field is included in the calculations using a macrocell approach~\cite{vampire} with a cell size of 1 nm. 
The used parameters are reported in Table~\ref{table:atomstic_parameters}.
\begin{table}[!htb]
\caption{Simulation parameters for the investigated systems.}
\begin{ruledtabular}
\begin{tabular}{cccc}
  			&   CoFeB(@interface)     			& CoFeB(bulk) 			& Unit            \\
  \cline{1-4}
  \smJij    & $1.547 \times 10^{-20}$  	& $7.735\times 10^{-21}$& J link$^{-1}$\\ 
  \smmu     & 1.60		               	& 1.60 					& \muB\\
  \smKu     & $1.35\times 10^{-22}$		& 0.0					& J atom$^{-1}$ \\
\end{tabular}
\end{ruledtabular}
\label{table:atomstic_parameters}
\end{table}

In the hysteresis loop calculations we use a critical damping and calculate a complete hysteresis cycle over 20 ns with an effective field rate of 0.3 T ns$^{-1}$ to minimize the effects of enhanced coercivity caused by fast field sweep rates.
%
\bibliographystyle{apsrev4-1}
\bibliography{Rev-mech}
%
\begin{acknowledgments}
This work was supported by the Samsung Global MRAM Innovation Program. The authors would like to thank Sara Majetich and Oksana Chubykalo-Fesenko for helpful discussions. This work made use of the facilities of N8 HPC Centre of Excellence, provided and funded by the N8 consortium and EPSRC (Grant No.EP/K000225/1).
\end{acknowledgments}
\section*{Author contributions}
A.M. performed the atomistic simulations, data analysis and produced the figures. All authors contributed to the design and direction for the study and to writing the manuscript. 
\section*{Conflicts of interest}
The authors declare no conflicts of interest.
%
 \pagebreak
 \newpage
\setcounter{equation}{0}
\setcounter{figure}{0}
 \renewcommand{\theequation}{Supplementari Eq. \arabic{equation}}
 \renewcommand{\thefigure}{Supplemntary Fig. \arabic{figure}}
%
\section*{Supplementary methods: Dependence of coercivity on magnetic moment}
\begin{figure}[!hb]
	\centering
	\includegraphics[width=1.0\columnwidth]{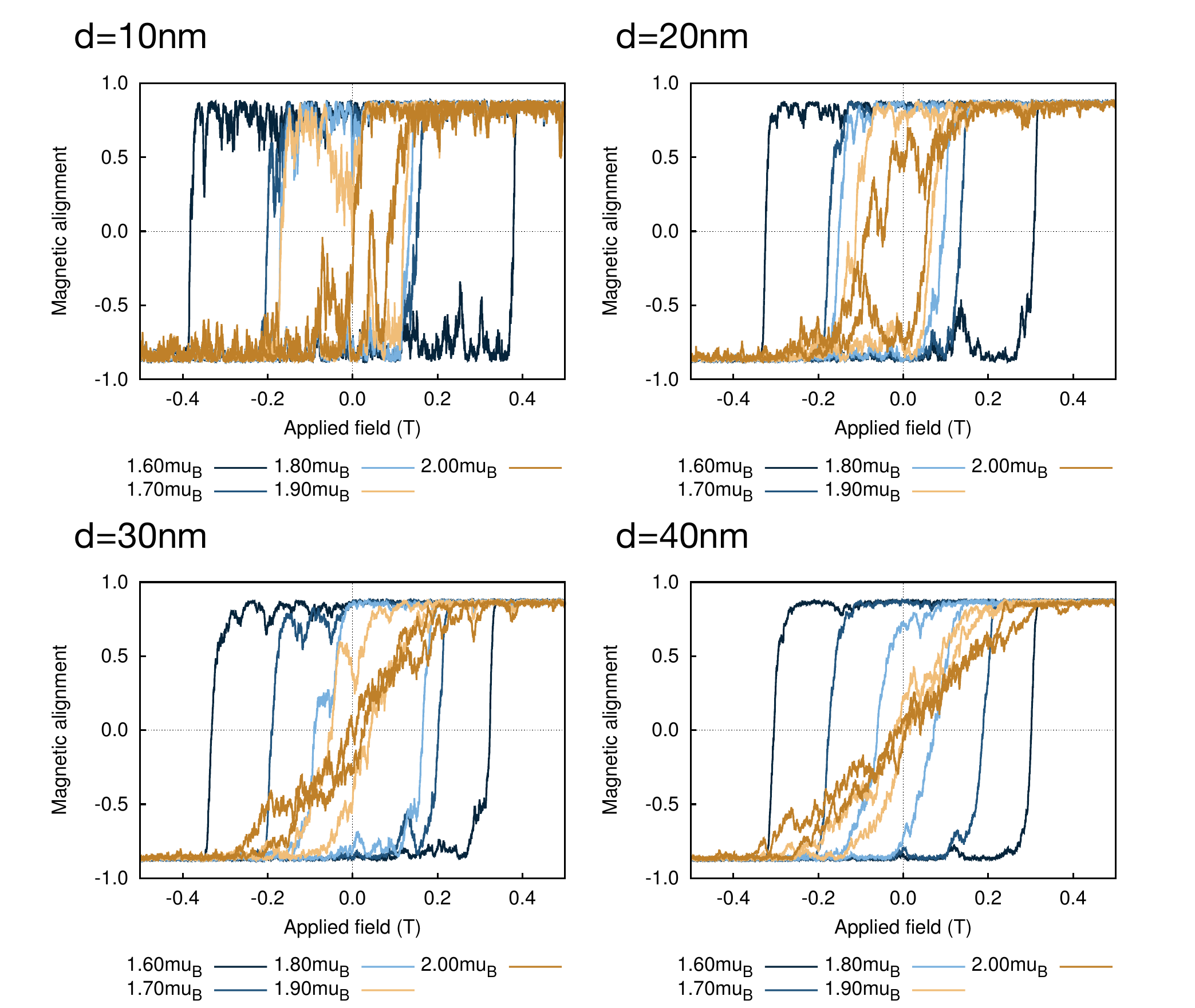}
	\caption{Hysteresis loops performed at room temperature for nanodots of thickness 1.3~nm and diameter 10, 20, 30, 40~nm as function of atomic magnetic moment (\smmu), from 1.6 to 2.0~\muB. Colours refer to different \smmu values.}
	\label{fig:SFigure1}
\end{figure}

\ref{fig:SFigure1} shows hysteresis loops performed at room temperature for nanodots of thickness 1.3~nm at room for various diameters as a function of the magnetic moment. Loops for systems characterized by a magnetic moment of 2.0~\muB show that the magnetisation lies in-plane. For smaller values an out-of-plane character can be observed. In this range, a magnetic moment of 1.6~\muB gives loops that are square and stable as seen in experiments.
\section*{Supplementary note 1: Effect of temperature on coercivity for small dots}
\begin{figure}[!tb]
	\centering
	\includegraphics[width=0.75\columnwidth]{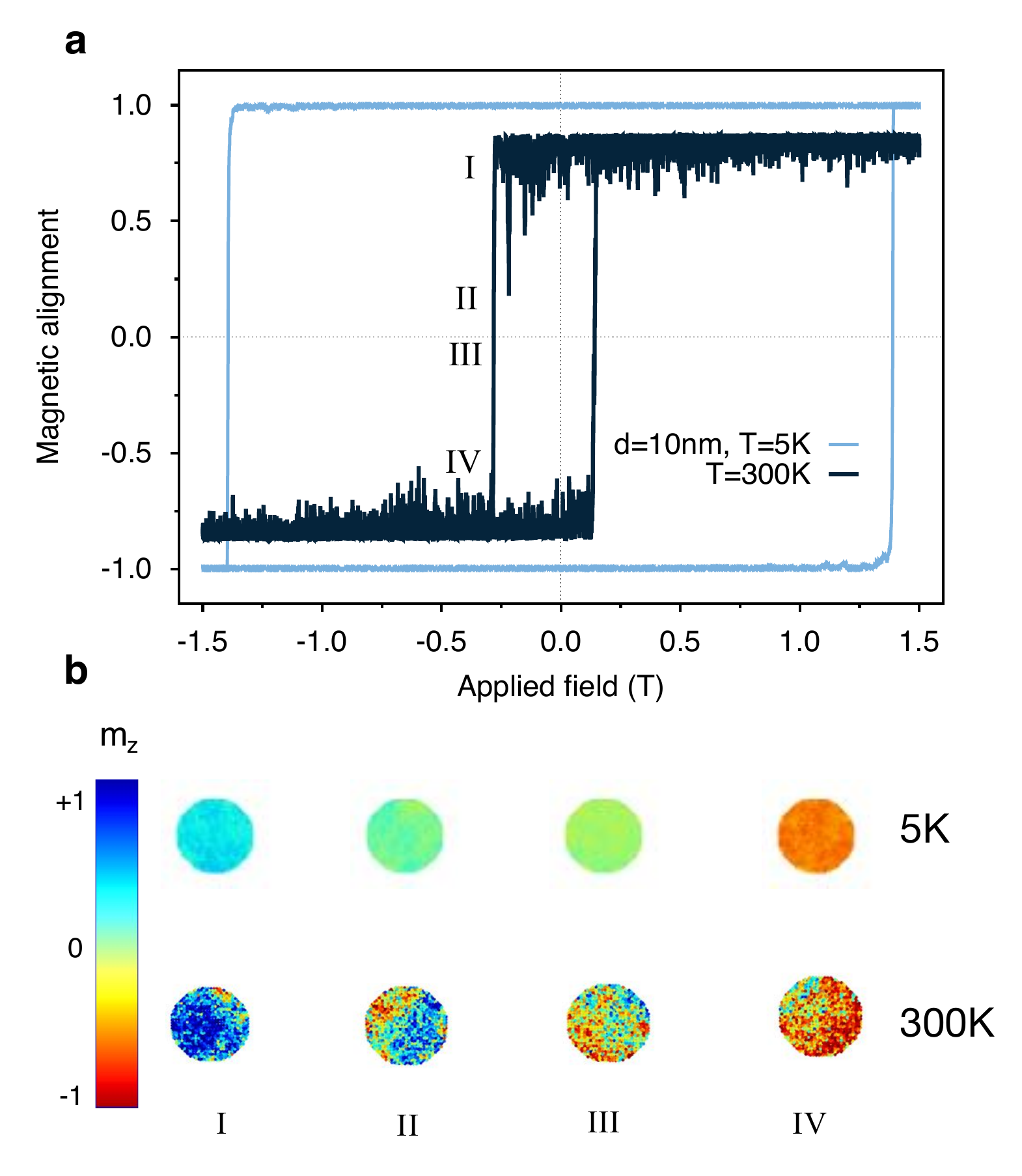}
	\caption{(a) Typical simulated easy-axis hysteresis loop for 1~nm thick, 10~nm diameter nanodot at temperatures of 5~K and 300~K. The data show a large reduction in the coercivity for elevated temperatures due to increased thermal fluctuations, indicating a change in the magnetic reversal mechanism. (c) Snapshots of magnetisation reversal at 5 and 300~K for a disk of diameter 10~nm and thickness 1~nm. I and IV refer to the top and bottom shoulder of M/\Mags vs H curve, respectively. II and III are configurations just before and after the switching, respectively. The colour scheme represents the magnetisation along the easy axis direction (z).}
	\label{fig:SFigure2}
\end{figure}

\ref{fig:SFigure2} (a) shows the hysteresis loops at 5~K and 300~K for 10~nm dots. A net reduction of the coercivity at room temperature emerges compared with 50~nm dots as well as larger thermal contribution which results in a pronounced asymmetry of the two branches and increase in the noise. From the analysis of the spin configurations in \ref{fig:SFigure2} (b) the reversal appears dominated by fluctuation of the magnetisation due to the small volume. At low temperature the system exhibits high coercivity and the reversal mechanism becomes quasi coherent. As observed for larger dots, the hysteresis loops are symmetric and the thermal fluctuations of the magnetisation become negligible. We point out that even at low temperature the coercivity is expected to decrease as the diameter of the dots is further decreased, due to loss of thermal stability.
\section*{Supplementary note 2: Dependence of average coercive field on diameter for large dots}
\begin{figure}[!tb]
	\centering
	\includegraphics[width=0.75\columnwidth]{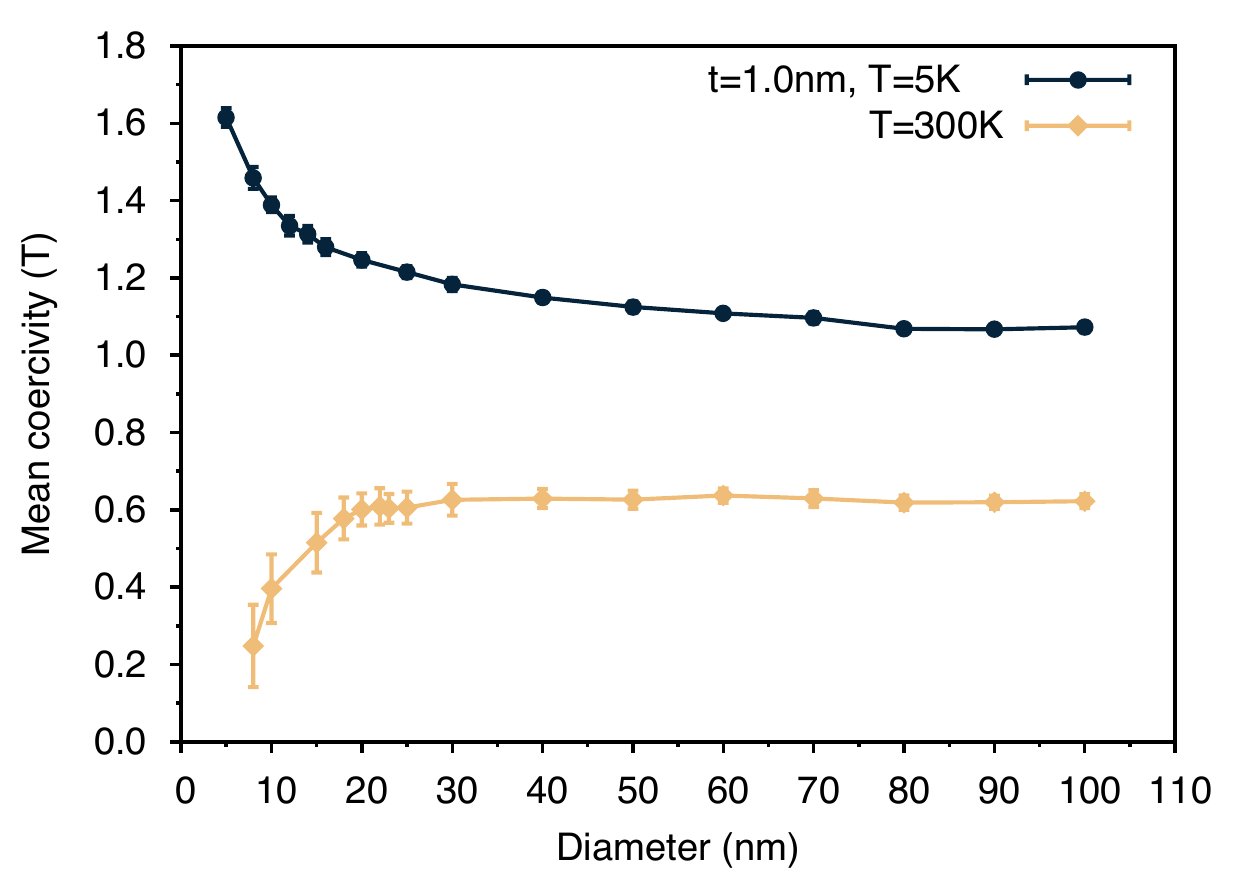}
	\caption{Mean coercivity as function of disk diameter for nanodots of thickness of 1.0~nm at 5 (black dots) and 300~K (yellow diamonds). Error bars show the standard deviation of the statistical distribution.}
	\label{fig:SFigure3}
\end{figure}

We calculate the average coercive field for large dots (70~nm) at 300~K and 5~K as function of diameter. The result is shown in \ref{fig:SFigure3}. In the low temperature case the average coercive field tends to an asymptotic value for dots larger than 60-70~nm and the reversal mechanism is non-uniform when the demagnetisation field of the nanodots approaches the saturation limit of a thin film yielding to a constant nucleation field. At room temperature the average coercivity reaches a limit value for nanodots diameters larger than 20~nm, as discussed in relation to \ref{fig:SFD}.
\section*{Supplementary note 3: Magnetisation reversal in MTJ}
\begin{figure}[!tb]
	\centering
	\includegraphics[width=1.0\columnwidth]{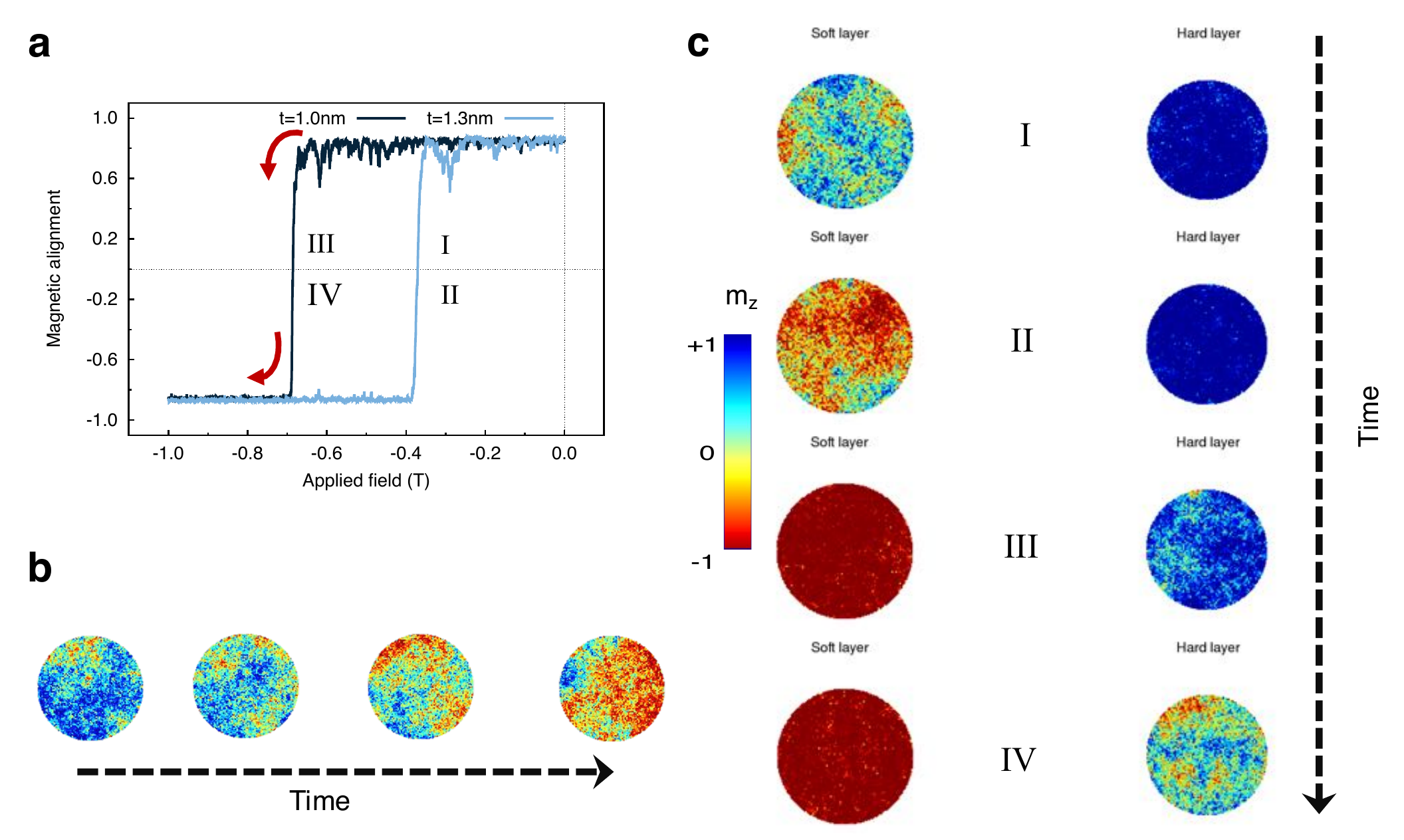}
	\caption{(a) Part of major loop from which the snapshots are taken. The roman numbers represent the field points at which the snapshots are taken. (b) Snapshots of magnetisation reversal at 300~K for a nanodot of diameter 20~nm and thickness 1.0~nm. (c) Snapshots of magnetisation reversal at 300 K for an MTJ of diameter 20~nm during a major loop. Left and right dots represent the free and pinned layer, respectively and roman numbers refer to the field points in (a). The colour scheme represents the magnetisation along the easy axis direction (orthogonal to the dot).}
	\label{fig:SFigure4}
\end{figure}

\begin{figure}[!tb]
	\centering
	\includegraphics[width=1.0\columnwidth]{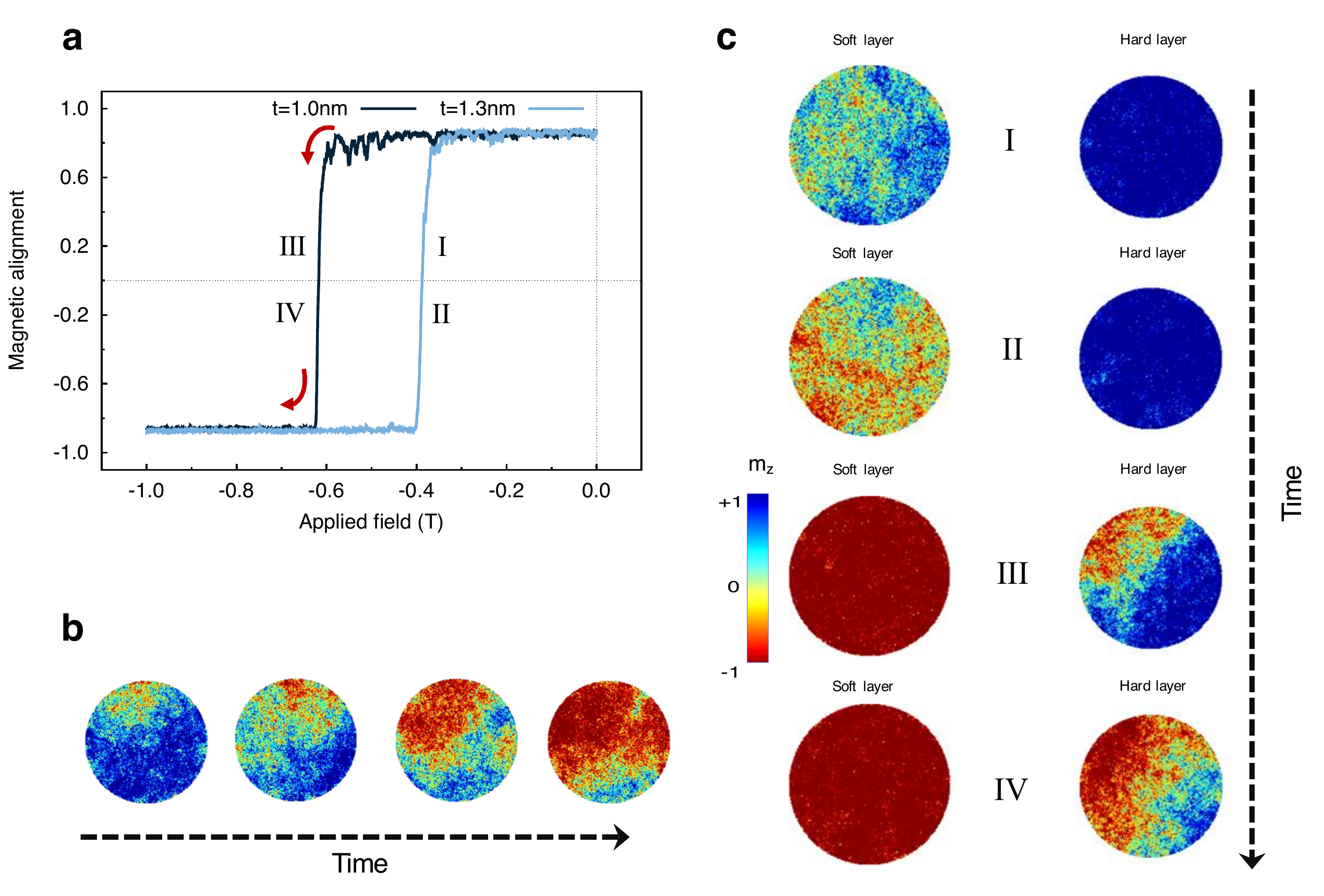}
	\caption{(a) Part of major loop from which the snapshots are taken. The roman numbers represent the field points at which the snapshots are taken. (b) Snapshots of magnetisation reversal at 300~K for a nanodot of diameter 30~nm and thickness 1.0~nm. (c) Snapshots of magnetisation reversal at 300~K for an MTJ of diameter 30~nm during a major loop. Left and right dots represent the free and pinned layer, respectively and roman numbers refer to the field points in (a). The colour scheme represents the magnetisation along the easy axis direction (z).}
	\label{fig:SFigure5}
\end{figure}

We have calculated the magnetic properties of a MTJ. A branch of a major hysteresis loop is shown in \ref{fig:SFigure4} (a), where both the free (soft) and pinned (hard) layers switch their magnetisation. It can be seen that the reversal modes of both layers show the same feature and do not reverse independently. This can be seen by comparing the magnetisation reversal configurations with the switching of the individual layers. \ref{fig:SFigure4} (b) presents the snapshots of the switching of the magnetisation for a nanodot of the same diameter and thickness 1.0~nm. From the comparison, we can observe as the mechanism of the reversal is the same: the magnetisation reversal occurs via thermal activation, therefore confirming that in our simulation the reversal mechanism of each single layer constituting the MTJ is not affected by the stacking. In \ref{fig:SFigure5} (a) the same results obtained for a MTJ of diameter 30~nm are presented. Conversely from the previous case, in \ref{fig:SFigure5} (b) we clearly observe edge nucleation, in agreement with the analysis proposed for the single layers at room temperature. A comparison of the hysteresis branch presented in (a) show how for 30~nm the magnetisation results more thermally stable close to the nucleation field.
\section*{Supplementary note 4: Spin transfer torque switching in MTJs}
We simulate the spin transfer torque switching of a MTJ with diameter 40~nm at room temperature under the application of a current density of $1\times 10^{11} \mathrm{A}\,\mathrm{m}^{-2}$.
The ferromagnetic layers are described with low damping ($\alpha$=0.003) and the spin torque switching is modelled such that it is originated at the interface CoFeB(Free layer)/MgO. We analyse the time evolution of the spin configurations during the switching. A clear domain wall edge nucleation and subsequent propagation emerges from the snapshots. The stochastic nature of the thermal fluctuations is responsible for the edge nucleation allowing nucleation sites localized at the edge of the system. 
\pagebreak
\section*{Supplementary figures}

\end{document}